\documentclass{article}
\usepackage{spconf,amsmath,graphicx,hyperref,booktabs,multicol,multirow}
\usepackage{enumitem}
\usepackage{pifont}
\usepackage{cite}
\newcommand{\cmark}{\ding{51}}%
\newcommand{\xmark}{\ding{55}}%
\usepackage[caption=false, position=bottom]{subfig}

\title{Arti-6: Towards six-dimensional articulatory speech encoding}

\name{Jihwan Lee$^1$, Sean Foley$^{1,2}$, Thanathai Lertpetchpun$^1$, Kevin Huang$^1$, Yoonjeong Lee$^1$, \\\em{Tiantian Feng$^1$, Louis Goldstein$^2$, Dani Byrd$^2$, Shrikanth Narayanan$^{1,2}$}}

\address{$^1$Signal Analysis and Interpretation Lab, University of Southern California \\
$^2$Department of Linguistics, University of Southern California}
\begin{document}
%
\maketitle
\begin{abstract}
We propose \textsc{ARTI-6}, a compact six-dimensional articulatory speech encoding framework derived from real-time MRI data that captures crucial vocal tract regions including the velum, tongue root, and larynx.
\textsc{ARTI-6} consists of three components: (1) a six-dimensional articulatory feature set representing key regions of the vocal tract; (2) an articulatory inversion model, which predicts articulatory features from speech acoustics leveraging speech foundation models, achieving a prediction correlation of 0.87; and (3) an articulatory synthesis model, which reconstructs intelligible speech directly from articulatory features, showing that even a low-dimensional representation can generate natural-sounding speech. Together, \textsc{ARTI-6} provides an interpretable, computationally efficient, and physiologically grounded framework for advancing articulatory inversion, synthesis, and broader speech technology applications.
The source code and speech samples are publicly available.\footnote{\url{https://github.com/lee-jhwn/arti-6}}
\end{abstract}
\begin{keywords}
Articulatory Inversion, Articulatory Synthesis, Real-time MRI, Articulatory Features, Speech Encoding
\end{keywords}
\vspace{-5mm}
\section{Introduction}
\label{sec:intro}
\vspace{-3mm}
Speech production involves coordinated, goal-directed  movements of the vocal tract articulators to achieve a sequence of phonological constriction tasks, or gestures~\cite{browman1992articulatory}. These articulatory dynamics shape the resulting acoustic speech signal, but in a way that does not transparently reveal its articulatory source. \textbf{Articulatory inversion}, also referred to as speech inversion or acoustics-to-articulatory inversion, is the task of inferring these hidden articulatory movements from speech acoustics, and it has been pursued as a means to uncover this hidden layer of speech production~\cite{atal1978inversion, hogden1996accurate, ghosh2010generalized, lammert2013statistical, peter_aai_2023, cho2023evidence}. \textbf{Articulatory synthesis} reconstructs speech from articulatory features, offering a physiologically grounded pathway for speech generation~\cite{otani2023speech, wu22i_interspeech, cho2024coding, tabatabaee25b_interspeech}. Together, articulatory inversion and synthesis provide a unified framework for understanding human speech production and developing interpretable, controllable speech technologies~\cite{shi2024direct, huang25h_interspeech,mcghee25_interspeech}.


 A recent work, named SPARC~\cite{cho2024coding}, has reported promising performance on articulatory inversion and synthesis, but this model relies on articulatory data from Electromagnetic Articulography (EMA), which captures only a subset of vocal tract regions. It does not index actions of the velum, tongue root, and larynx, so it lacks explicit nasality and voicing information. To compensate for this limitation, however, the articulatory feature set in this work is augmented via the use of acoustic correlates (e.g., pitch, energy).

Articulatory data collected using real-time MRI (rtMRI) provides fuller coverage of the moving vocal tract, including the missing information from the EMA data, such as of the velum, tongue root, and larynx areas~\cite{narayanan2014real, sorensen2017database, hagedorn2019engineering, lim2021multispeaker}, and has been widely used 
in both engineering and scientific studies~\cite{toutios2019advances, hagedorn2019engineering, yue2024towards, shi2024direct, foley2025towards}. That said, a persistent challenge across modalities remains: both image-derived articulatory data and acoustic embeddings are often high-dimensional and opaque, while overly simplified features risk discarding essential information. This motivates the search for a compact, interpretable representation that captures key articulatory dynamics while remaining tractable for modeling.


In this work, we introduce ARTI-6, a six-dimensional articulatory speech encoding framework derived from knowledge-driven selection of key vocal tract regions critically involved in speech production captured with rtMRI.
We demonstrate that this compact representation supports both high-quality articulatory inversion and intelligible articulatory synthesis.

Beyond strong performance, \textsc{ARTI-6} offers interpretability within the vocal tract space used for speech. Physiologically grounded in articulatory kinematics and theoretically motivated by linguistic models of speech representation~\cite{browman1992articulatory}, its six regions capture both the physical dynamics and the cognitive structure of speech sounds. This dual grounding makes inferred articulatory features valuable for scientific research where direct articulatory data acquisition is challenging~\cite{lee25d_interspeech, huang25h_interspeech, chartier2018encoding}.
At the same time, the low dimensionality of \textsc{ARTI-6}, is far more compact than conventional intermediate speech features such as MFCCs or mel-spectrograms and recent deep-learning based latent speech representations that span hundreds of dimensions. This compactness brings computational efficiency and low latency, making \textsc{ARTI-6} a practical intermediate feature space for real-time speech applications. A comparison of \textsc{ARTI-6} with other intermediate speech feature sets is summarized in Table~\ref{tab:compare-model}.



\begin{figure*}[t]
  \centering
  \includegraphics[width=\linewidth]{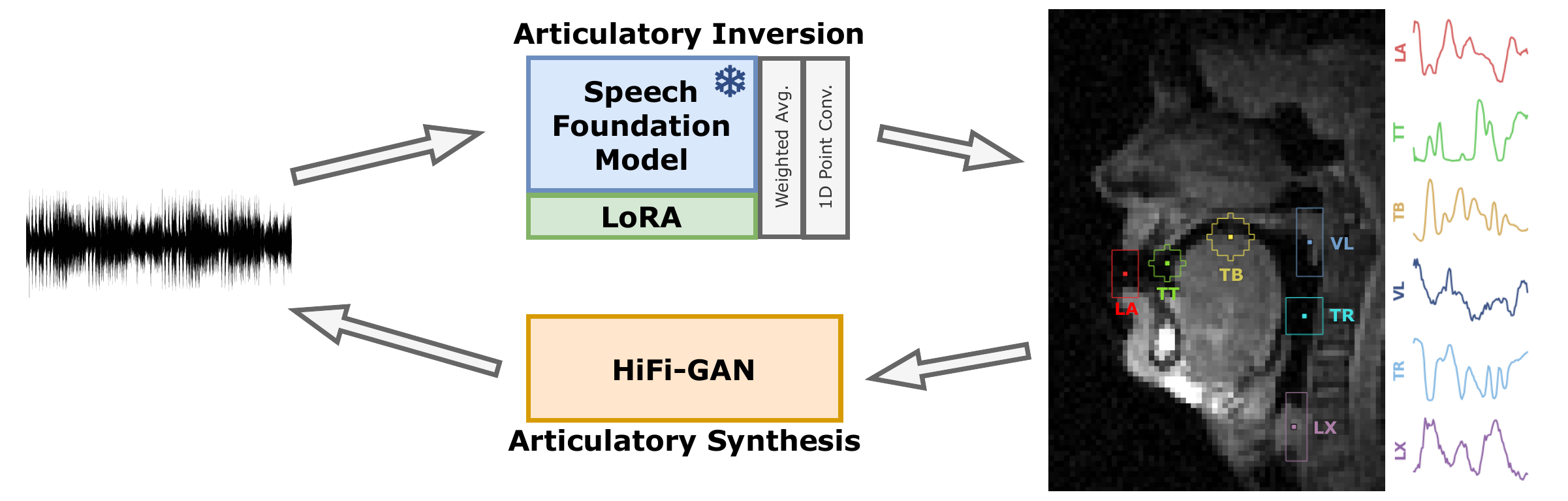}
  \vspace{-7mm}
  \caption{Overview of the \textsc{ARTI-6} framework. The articulatory inversion and synthesis models, and the six key regions of interest (ROIs) of the six-dimensional articulatory features (right): Lip Aperture (LA), Tongue Tip (TT), Tongue Body (TB), Velum (VL), Tongue Root (TR), and Larynx (LX).}
  \label{fig:emg-ema}
  \vspace{-5mm}
\end{figure*}

\section{Methods}
\label{sec:method}


\subsection{\textsc{ARTI-6}: Region Selection}
We choose six key regions of interest (ROIs) from the vocal tract~\cite{usc-lss} that are crucially involved in the achievement of speech production goals through constriction formation and release: Lip Aperture (LA), Tongue Tip (TT), Tongue Body (TB), Velum (VL), Tongue Root (TR), and Larynx (LX).
From each region, we use the image pixel intensity as a proxy measure for constriction degree of the vocal tract variables.
These choices are motivated by three assumptions: 1) the selected regions represent the minimal articulatory requirement for speech representation~\cite{browman1992articulatory}; 2) other articulatory kinematic points in rtMRI highly correlate with these representative regions; and 3) models can learn to infer the full articulatory dynamics from these representative regions.
The precise locations of the ROIs and the corresponding pixel intensity were extracted using the VocalTract ROI Toolbox~\cite{roitool}.

\begin{table*}[t]
    \vspace{-7 mm}
  \caption{Comparison of our proposed \textsc{ARTI-6} with other intermediate speech feature types.}
  \label{tab:compare-model}
  \centering
  \begin{tabular}{ccccccc}
    \toprule
    \multirow{2}{*}{\textbf{Intermediate Feature}} & \multirow{2}{*}{Dimension Size} & \multirow{1}{*}{Articulatory} & \multirow{2}{*}{Knowledge-based} & \multicolumn{3}{c}{Direct coverage of ...}  \\ 
     & & Intrepretablility & & Velum & Tongue Root &Larynx \\
    \toprule
    MFCC & $10$-$20$ & \xmark & \cmark Speech Perception & \xmark & \xmark & \xmark  \\
Mel-Spectrogram & $80$ & \xmark & \cmark Speech Perception & \xmark & \xmark & \xmark  \\
Latent Spaces~\cite{baevski2020wav2vec, chen2022wavlm, hsu2021hubert} & large ($200+$) & \xmark & \xmark & \xmark & \xmark & \xmark\\

 EMA+pitch+loudness~\cite{cho2024coding}&  $14$ & \cmark & \cmark Speech Production & \xmark & \xmark & \xmark\\
 \midrule
  \textsc{ARTI-6} (Ours)&  $6$ & \cmark & \cmark Speech Production& \cmark & \cmark & \cmark \\

    \bottomrule
  \end{tabular}
      \vspace{-5mm}
\end{table*}

\subsection{Articulatory Inversion}
\label{sec:articulatory-inv}

We base the articulatory inversion modeling on speech foundation models such as \texttt{Whisper}~\cite{radford2023robust}, \texttt{WavLM}~\cite{chen2022wavlm}, \texttt{HuBERT}~\cite{hsu2021hubert}, and \texttt{Wav2Vec2}~\cite{baevski2020wav2vec}.
\texttt{Whisper-Large} is a multilingual model originally trained for automatic speech recognition, language identification, and voice activity detection, and the other models are trained with self-supervised learning methods using contrastive, predictive, or clustering-based objectives. Our articulatory inversion model is fine-tuned using \texttt{LoRA}~\cite{hu2022lora}, and the trainable low-rank matrices are inserted in the feed-forward layers in each Transformer~\cite{transformer} layer. Moreover, we introduce a learnable linear layer that combines the hidden states from both convolutional and Transformer layers. The aggregated output is then fed through 1D-pointwise convolutional layers followed by a projection layer that maps to the six-dimensional articulatory features. 


\subsection{Articulatory Synthesis}
\label{sec:articulatory-syn}

We base our articulatory synthesis model on the \texttt{HiFi-GAN v1}~\cite{kong2020hifi} architecture. We use upsample rates and kernel sizes of $[8,5,4,2]$ and $[16,10,8,4]$, respectively. The segment and hop sizes are set as $8000$ and $320$, to match the output audio sampling rate of $16$ kHz. The rest of the model configurations are identical with the original version. We also condition the articulatory synthesis model on speaker embeddings extracted from \texttt{ECAPA-TDNN}\footnote{\url{https://huggingface.co/speechbrain/spkrec-ecapa-voxceleb}}~\cite{desplanques2020ecapa}. We adopt the similar assumption as in \cite{cho2024coding},
that the articulatory synthesis model is able to capture the individual speaker characteristic of the vocal tract from the conditioned speaker embedding.

\section{Experiments}
\label{sec:experiments}
\subsection{Datasets}
For articulatory inversion, we used the USC Long Single-Speaker (LSS) dataset ~\cite{usc-lss}, the longest rtMRI articulatory dataset from a single speaker. It has $684$ utterances totaling $54$ minutes of speech from a single American English speaker. This includes $37$ minutes of read speech and $17$ minutes of spontaneous speech. We adopt the identical train, dev, and test set split as in \cite{usc-lss}, containing $581$, $34$, and $69$ utterances, respectively. The dataset contains audio sampled at $16$ kHz and rtMRI video reconstructed at $99$ frames/sec. Hand-corrected phoneme alignments are also included in the dataset. 

For articulatory synthesis, we augmented with the LibriTTS-R~\cite{librittsr} dataset, as the LSS dataset solely is not sufficient to train a deep-learning based speech synthesizer. Using the articulatory inversion model of the \texttt{WavLM} backbone, we created pairs of predicted articulatory features and speech acoustics and used them to train the articulatory synthesis model. We used \texttt{train-960}, \texttt{dev-clean}, and \texttt{test-clean} for train, dev, and test set, respectively.
All reported results are from the test set, unless otherwise mentioned. 

\subsection{Pre-processing}
We leveraged the recent speech restoration model \texttt{Miipher}\footnote{\url{https://github.com/Wataru-Nakata/miipher}} \cite{koizumi2023miipher} to remove background noise created by the MRI device from the speech acoustics. The speech restoration model takes text and noisy speech as input and outputs an enhanced (or restored) version of the speech. 
The extracted articulatory features from rtMRI videos were resampled from the original $99$ Hz frame rate to $50$ Hz frame rate to match the frame rate of the latent space of the speech foundation models.

\begin{table}[t]
  \caption{Articulation inversion performance measured by the Pearson correlation between the predicted and target articulatory features.}
  \label{tab:invres}
  \centering
  \begin{tabular}{lcc}
    \toprule
    \multirow{2}{*}{\textbf{Speech Foundation Model}}&  \multicolumn{2}{c}{\textbf{\centering Prediction Correlation}}\\
     & dev & test\\
    \toprule
Wav2Vec2-XLSR~\cite{baevski2020wav2vec}
    &  $0.637$ & $0.615$ \\
Whisper-Large~\cite{radford2023robust} 
    & $0.864$ & $0.844$ \\

    HuBERT~\cite{hsu2021hubert}
    & $0.890$ & $0.872$ \\

       WavLM-Large~\cite{chen2022wavlm}
    & $0.892$ & $0.872$ \\
    \bottomrule
  \end{tabular}
\end{table}

\subsection{Training}
For the articulatory inversion model, the L2 loss was adopted as the fine-tuning objective, optimized by the AdamW~\cite{loshchilov2018decoupled} optimizer with a batch size of $4$.
For the articulatory synthesis model, we adopted the same training objectives from \cite{kong2020hifi}: the GAN loss~\cite{mao2017least}, the reconstruction loss, and the feature matching loss. It was trained for $1.2M$ steps with a batch size of $16$.
Each experiment was trained on a single NVIDIA L40S GPU.
Refer to the source code for full details.

\section{Results and Discussion}
\label{sec:results}

\begin{figure}[t]
  \includegraphics[width=\linewidth]{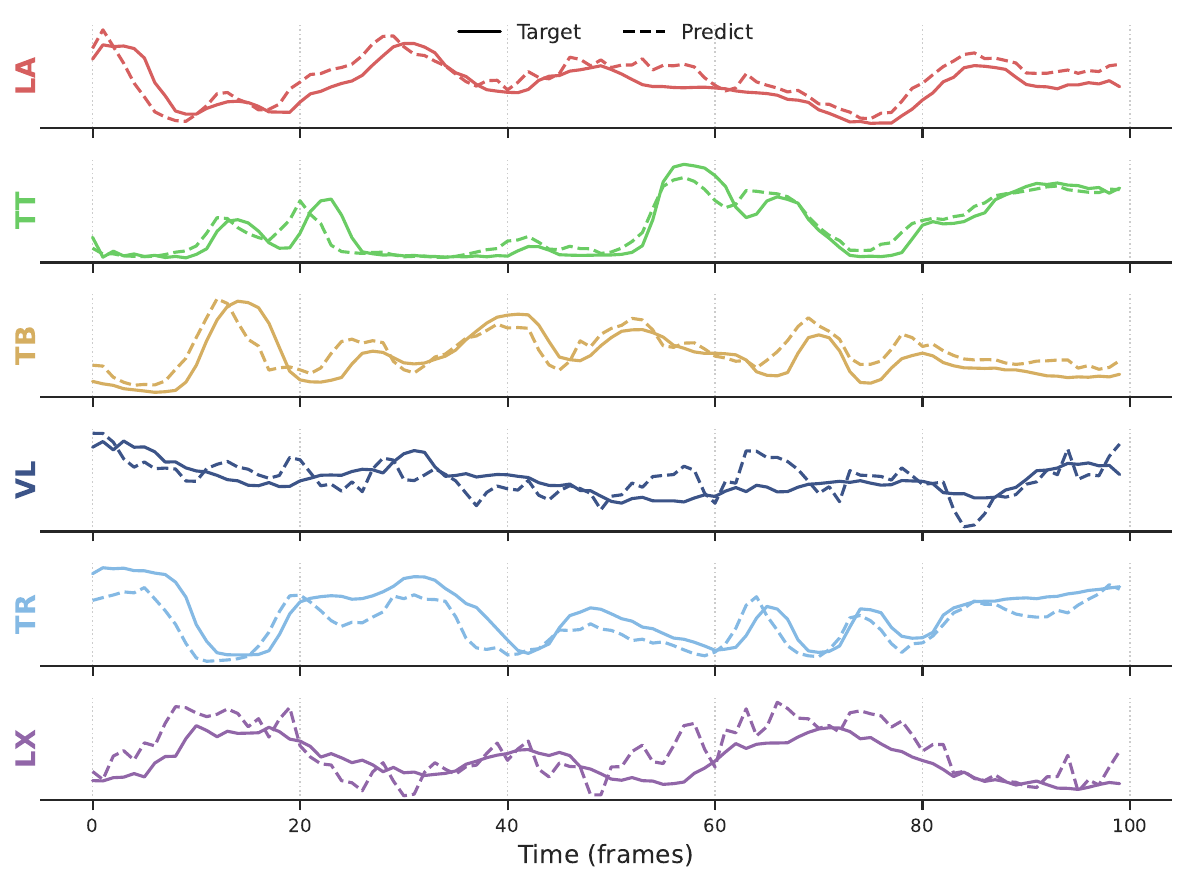}
  \caption{An example utterance of predicted and target articulatory features of \textsc{ARTI-6}.
  High prediction accuracy is achieved for the lip and tongue regions, whereas performance is comparatively lower for the velum and larynx regions.}
  \label{fig:pred}
  \vspace{-3mm}
\end{figure}

\begin{table*}[ht]
  \caption{Evaluation on synthesized speech waveforms. Word Error Rate (WER), Character Error Rate (CER), UTMOS, and Mean Opinion Score (MOS) with 95\% confidence intervals.}
  \label{tab:subchannel}
  \centering
  \begin{tabular}{l|c|c|c|c|c}
    \toprule
    \textbf{Intermediate Feature}&  \textbf{Dim} &
    \textbf{WER $\downarrow$} & \textbf{CER $\downarrow$} &\textbf{UTMOS $\uparrow$}& \textbf{MOS $\uparrow$} \\ 
    \toprule
    Ground-truth
    &  - & $0.034 \pm 0.003$ & $0.021 \pm 0.006$ & $4.196 \pm 0.004$ & $4.604 \pm 0.081$\\
    Mel-spectrogram~\cite{kong2020hifi}
    & $80$ & $0.038 \pm 0.003$ & $0.025 \pm 0.006$ & $3.975 \pm 0.005$ &$4.447 \pm 0.103$\\

    EMA+pitch+loudness~\cite{cho2024coding}
    & $14$ & $0.048 \pm 0.003$ & $0.031 \pm 0.006$ & $4.164 \pm 0.004$ &$4.265 \pm 0.121$\\

        \textsc{ARTI-6} (Ours)
    & $6$ & $0.125 \pm 0.004$ & $0.074 \pm 0.006$ & $3.840 \pm 0.005$ & $3.947 \pm 0.135$\\

    \bottomrule
  \end{tabular}
  \vspace{-3mm}
\end{table*}

\subsection{Articulatory Inversion}
We evaluate the articulatory inversion performance by measuring the Pearson correlation between the target and the predicted articulatory features. Among the four speech foundation models we experimented with, we observe the highest performance when \texttt{WavLM}~\cite{chen2022wavlm} or \texttt{HuBERT}~\cite{hsu2021hubert} is used as the backbone, achieving a prediction correlation of $0.872$, as shown in Table~\ref{tab:invres}.  Figure~\ref{fig:pred} shows an example of predicted and target articulatory features.

\begin{figure}[t]
  \includegraphics[width=\linewidth]{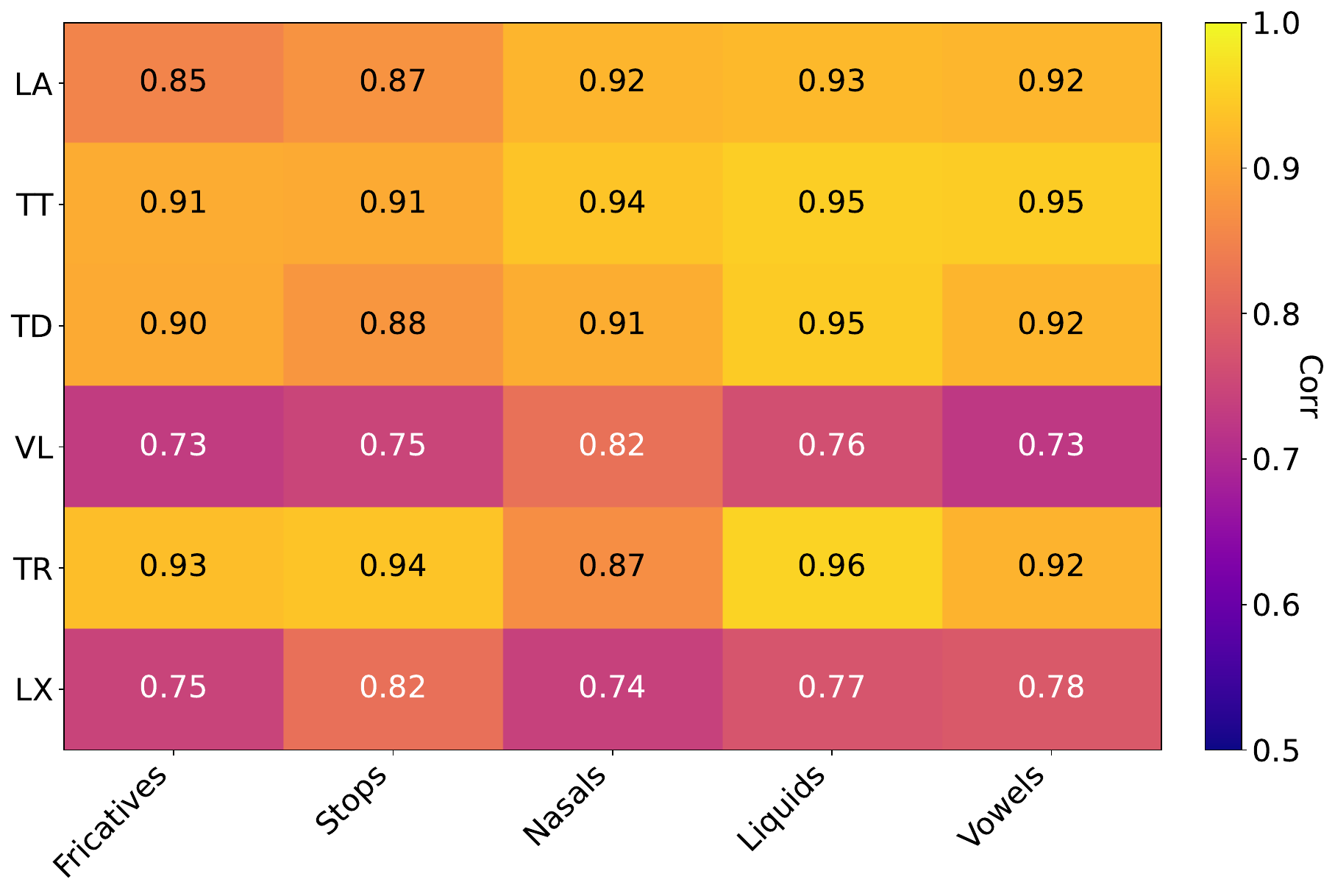}
  \vspace{-7mm}
  \caption{Heatmap of prediction correlations with respect to the vocal tract ROIs and phonetic manner categories.}
  \label{fig:phn}
  \vspace{-5mm}
\end{figure}

We next analyze the articulatory inversion performance at the phonemic level with respect to each ROI of the vocal tract, as shown in Figure~\ref{fig:phn}. The phoneme alignments provided in \cite{usc-lss} were used to locate each phoneme segment. To compensate for the  influence of  duration on the prediction correlation, the Fisher $z$-transformation was applied to the raw correlation, followed by a duration-based scaling.
Greater prediction performance is observed for the lip and tongue regions, compared to the velum and larynx regions, as shown in Figure~\ref{fig:phn}. For the velum, this may be attributable to noise in the pixel intensity measure during velum closure. During sequences of oral sounds the velum typically stays closed, though there may still be fluctuations in pixel intensity within the velum ROI unrelated to actively controlled velum movement. Similarly, for the larynx the midsagittal orientation of the rtMRI video may be limited in capturing the larynx's detailed internal geometry dynamics.
This phonemic analysis provides the model characteristics of the current approach and points to potential future directions for optimizing both the feature set and the modeling strategy.

\subsection{Articulatory Synthesis}

We evaluate the quality of synthesized speech using both objective and subjective metrics. For the objective evaluation, we measure Word Error Rate (WER), Character Error Rate (CER), and UTMOS~\cite{saeki22c_interspeech}. UTMOS~\cite{saeki22c_interspeech} is a proposed metric that automatically predicts Mean Opinion Scores (MOS) from speech acoustics. The WER and CER are measured by \texttt{Whisper-large-v3}. The reported values are measured from all of the speech samples from \texttt{test-clean} in LibriTTS-R~\cite{librittsr}. For the subjective metrics, we measure naturalness MOS from $17$ human evaluation responses. The evaluators are speech researchers who are native or advanced-level speakers of English, currently residing in the USA. The evaluators were asked to rate the naturalness of each speech sample on the scale of \textit{[1-not at all natural, 2-slightly natural, 3-moderately natural, 4-quite natural, and 5-extremely natural]}. Only a subset of $40$ utterances from the same test set were evaluated in the human evaluation protocol.

Although the overall quality of speech generated with \textsc{ARTI-6} is lower than that of high-dimensional approaches, this outcome is not surprising given its extremely compact representation. What is noteworthy, however, is that even with only six articulatory dimensions, the reconstructed speech remains acceptably intelligible, as indicated by WER and CER around $12\%$ and $7\%$, respectively. Furthermore, both UTMOS and MOS results suggest that the reconstructed speech samples are \textit{quite natural}.
We encourage readers to listen to the speech samples on the sample page. This demonstrates the potential of \textsc{ARTI-6} as a lightweight, interpretable, and physiologically grounded representation for speech tasks. This finding highlights a promising direction for developing interpretable and efficient speech technologies. 

\vspace{-3mm}
\subsection{Potential Applications}

\textsc{ARTI-6} has potential applications in both the scientific and engineering domains. In scientific research for example, it can be valuable in experimental settings for which it is challenging to obtain concurrent recordings of articulatory kinematics with other bio-signal modalities (e.g., brain/muscle signals). In such cases, estimating articulatory features from speech acoustics can provide insights into speech–brain/muscle action relationships, as demonstrated in \cite{chartier2018encoding,lee25d_interspeech}. Further, compared to EMA-based representations, \textsc{ARTI-6} additionally captures articulatory regions such as the velum, tongue root, and larynx, offering a fuller coverage of the vocal tract articulatory space.
For engineering applications, \textsc{ARTI-6} can be advantageous in settings where low latency or computational efficiency is desired, even with some quality degradation tradeoff,
such as on-device processing, phone-line transmission, or streaming/real-time speech manipulation.



\vspace{-3mm}
\section{Conclusion and Future Work}
\label{sec:conclusion}

We introduce a six-dimensional articulatory feature representation framework, \textsc{ARTI-6}, that supports both articulatory inversion and articulatory synthesis. 
By grounding the representation in knowledge-driven selection of articulatory regions, ARTI-6 balances interpretability, low-dimensionality, and computational efficiency.

Our current work is limited to a single-speaker articulatory space. In future work, we aim to extend the framework to multi-speaker articulatory spaces, enabling broader generalization across speakers without relying on speaker-specific conditioning \cite{Ghosh2011AutomaticSpeechrecognitionusing}. In conjunction with this, we will further refine the derived tongue variables to be more comparable across speakers. We also plan to explore optimization of articulatory regions and proxy measures to further refine the image-derived representation. We will also explore optimizing the feature sets consisting of articulatory and acoustic features, to best complement each other.

\footnotesize
\section{Acknowledgment}
\vspace{-2.5mm}
This work was supported by the US4 National Science Foundation (IIS-2311676, BCS-2240349) and by the Office of the Director of National Intelligence (ODNI), Intelligence Advanced Research Projects Activity (IARPA), via the ARTS Program under contract D2023-2308110001. The views and conclusions contained herein are those of the authors and should not be interpreted as necessarily representing the official policies, either expressed or implied, of ODNI, IARPA, or the U.S. Government. The U.S. Government is authorized to reproduce and distribute reprints for governmental purposes notwithstanding any copyright annotation therein.
\vspace{-2mm}



\bibliographystyle{IEEEbib}
\footnotesize
\bibliography{strings,refs}

\end{document}